      [1999/12/01 v1.4c Il Nuovo Cimento]
\documentclass{cimento}

\usepackage{graphicx}  
\title{On the Origin of the Dark Gamma-Ray Bursts}
\author{A. M\'esz\'aros\from{ins:x}\ETC,
Z. Bagoly\from{ins:y},
S. Klose\from{ins:z},
F. Ryde\from{ins:u},
S. Larsson\from{ins:u},
L. G. Bal\'azs\from{ins:v},
I. Horv\'ath\from{ins:w},
L. Borgonovo\from{ins:u}
}
\instlist{\inst{ins:x} Astronomical Institute of the Charles University,
V Hole\v{s}ovi\v{c}k\'ach 2, CZ-180 00 Prague 8, Czech Republic
  \inst{ins:y} Laboratory for Information Technology,
E\"{o}tv\"{o}s University, H-1117 Budapest, P\'azm\'any P. s.  1./A, 
Hungary \inst{ins:z} Tautenburg Observatory, D-07778 Tautenburg,
Sternwarte 5, Germany \inst{ins:u} Stockholm Observatory, AlbaNova, SE-106 
91 Stockholm, Sweden \inst{ins:v} Konkoly Observatory, H-1525 Budapest, 
POB 67, Hungary \inst{ins:w} Department of Physics, Bolyai Military 
University, Budapest, Box-12, H-1456, Hungary  }

\PACSes{\PACSit{98.70}{98.80}}

\begin{document}

\maketitle

\begin{abstract}
The origin of dark bursts - i.e. that have no observed afterglows in X-ray,
optical/NIR and radio ranges - is unclear yet. Different possibilities -
instrumental biases, very high redshifts, extinction in the host ga\-laxies -
are discussed and shown to be important. On the other hand, the dark 
bursts should not form a new subgroup of long gamma-ray bursts  
themselves.
\end{abstract}

\section{Introduction}

In Jochen Greiner's list \cite{ref:gre},
up to June 30, 2004, there were collected 222 gamma-ray bursts (GRBs). Only 
71 objects (i.e. roughly the one-third fraction) have observed afterglows 
(AGs) at lower energies. In the following, the remaining two-thirds are 
defined as the fraction of the dark bursts. (Note that there is no 
unambigous definition of the dark burst itself, and usually
only that GRBs are dark, which have no optical AGs \cite{ref:jak}.
For our purpose the best definition is to take a GRB 
dark when there is no observed AG (regardless in which band) and 
also no redshift is measured.)

In accordance with \cite{ref:fyn} there can exist
four possible explanations for this phenomenon:

I. the observational biases play a role;

II. a large fraction of bursts is at high redshifts;

III. a large fraction of bursts is obscured by the interstellar matter in 
host galaxies;

IV. the darkness is intrinsic and some bursts really have no AGs.\\
Obviously, if only the first three possibilities were occurring, then any GRB 
would still have AG; i.e. for any GRB the AG would exist, 
but simply would not be 
detected. On the other hand, if the fourth possibility is also occurring, then 
there would be a special subclass of long GRBs having no AG.
The study of GRB 000330 \cite{ref:fyn} has shown that all
four eventualities were possible. \cite{ref:stra} and 
\cite{ref:lamb} give maximally a $\simeq (10-30)$ fraction for the IV-th 
explanation.

Our aim in this article is to search for more 
concrete conclusions concerning these four possibilities.

It is well-known that there are short and long
bursts; probably also intermediate ones (see \cite{ref:bal} and the 
references therein). Here we discuss only long
GRBs, because all GRBs in Greiner's list should belong to this subclass.
This also means that - if possibility IV were occurring - then the 
long subclass would be separated and there would be - in total - 4 
subclasses of GRBs. The most recent support for this separation of 
the long subgroup itself comes from \cite{ref:bor}. Nevertheless, 
repeating the same procedure with the gamma photometric 
redshifts of \cite{ref:bag}, we obtain 
no evident separation (Fig. 1). In this article we will give further 
arguments against the subclass separation of long GRBs.

\begin{figure}
\includegraphics[width= {1.0\columnwidth}, angle=0]{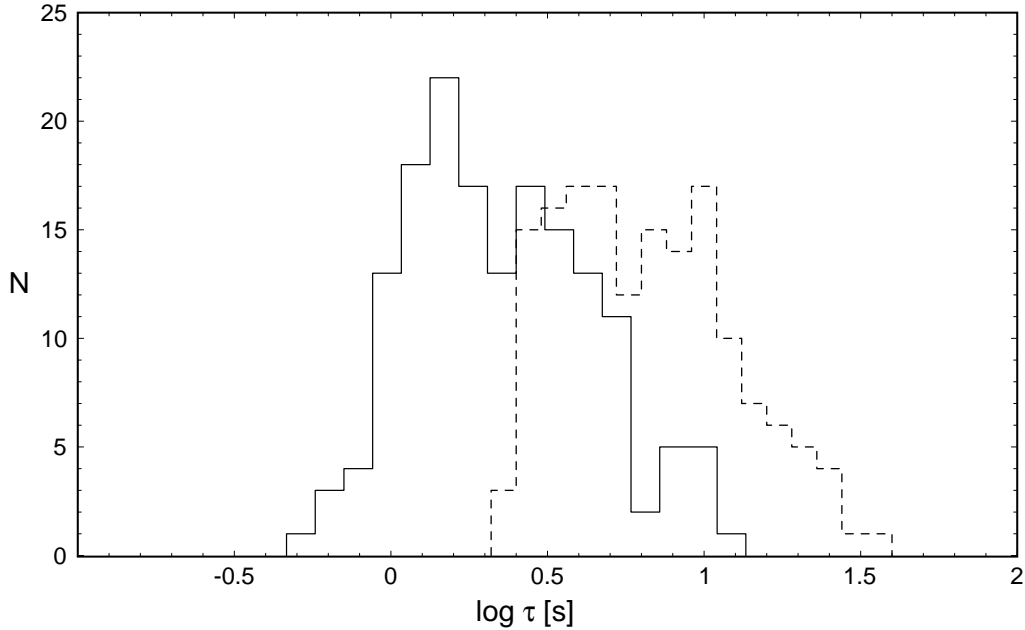}
\caption{Distribution of the 
autocorrelation function widths $\tau$ (for more details see
\cite{ref:bor}) for 160 GRBs. The redshifts were based on \cite{ref:bag}. The
logarithmic histograms of the widths are corrected (solid-line) or
not corrected (dashed-line) for the cosmological time dilation. In
both cases the distributions seem to be unimodal, in contradiction
with \cite{ref:bor}.}
\end{figure}

\section{The Method}

We will discuss the possibilities I-III.  Let us assume that all 
long GRBs have AGs, and the IV-th possibility is incorrect. If this is the 
case, then biases, GRBs at high redshifts, and absorption at the hosts 
must be, together, so effective that the two-third frac\-tion of GRBs 
must remain unseen due to these effects. We will show by statistical 
arguments that really such great amount of AGs may remain unseen by these 
effects. Then, simply, the fourth possibility is not needed.

The observed dark--non-dark separation is in the ratio $\sim 2:1$ at most.  
It is practically sure that some GRBs will have AGs, but these AGs remain 
unseen. As is shown by \cite{ref:me04}, there is a selection 
effect leading to the result that bright GRBs have 
more effectively observed AGs. This observational bias can 
be called a "brightness bias". Hence, there can be, in 
principle, three types of long GRBs:  GRBs
having observed AGs (non-dark GRBs) \& GRBs having AGs, but they remain 
unseen due to this bias 
("observationally dark" GRBs) \& GRBs having no AGs ("intrinsically dark" 
GRBs).  

\begin{table}
\caption{The redshifts ($z$) and peak-fluxes ($P_{256}$) of GRBs that have both 
BATSE
triggers and measured spectroscopic redshifts. The values are compiled
after \cite{ref:bor} using the Current BATSE Catalog \cite{ref:mee}.}
$$
\begin{array}{lllll}
\hline
 Burst & BATSE   & z &  P_{256}  &  Remark\\
       & trigger &   &  ph/(cm^2s)&   \\
\hline
970508 & 6225 & 0.835 &  1.17  & \\
970828 & 6350 & 0.9587 &   - & no\, P_{256}\, value \\
971214 & 6533 & 3.4127 &  2.32 & \\
980329 & 6665 & 3.97 & 13.28 & z\, upper\, limit\, only\\
980425 & 6707 & 0.0085 & 1.08  &  SN1998bw \\
980703 & 6891 & 0.966 &  2.59 & \\
990123 & 7343 & 1.60     &  16.63 & \\
990506 & 7549 & 1.3066&  22.16  & \\
990510 & 7560 & 1.619 & 10.20 & \\
991216 & 7906 & 1.020 &  15.17 & \\
000131 & 7975  & 4.5   & -   & no\, P_{256}\, value\\
 \hline
   \end{array}
$$
\end{table}

What is the ratio of the second type GRBs to the first type? It is   
sure that this ratio is non-zero, because both non-dark GRBs and     
observationally dark GRBs exist \cite{ref:me04}.
To answer this question consider the GRBs - observed by BATSE 
\cite{ref:mee} -  which have measured redshifts (either concrete
redshifts or at least upper limits) from the AG observations. In Table 1
these  GRBs are listed using \cite{ref:bor}. It is immediately remarkable 
that the $P_{256}$'s are unusually high.  For example, 5 GRBs have 
$P_{256} > 10$ {\em ph/(cm$^2$s)} among the 9 $P_{256}$  values. On the 
other hand, in the BATSE Catalog for $T_{90} > 10$ s only 9\% of GRBs 
fulfil the condition $P_{256}
> 10$ $ph/(cm^2s)$. In \cite{ref:me04} it is shown that this is
{\it not} chance: the brighter GRBs in gamma-ray range can be observed
more easily at lower energies.

In the BATSE Catalog 9\% of long GRBs are brighter than 10 {\em ph/(cm$^2$s)} 
in the $P_{256}$ peak-flux. Hence, if there were no observational 
biases, and if all GRBs were followed by AGs, one would detect roughly 10 
times more AGs for GRBs with peak-flux below 10 {\em ph/(cm$^2$s)} than 
above this level. Because we observe 5 AGs above the peak-flux level, we 
should see a further $\sim 50$ AGs below this $P_{256} = 10 ${\em 
ph/(cm$^2$s)} limit. Instead of 50 GRBs, 4 ones have been
detected. Instead of the 9 GRBs collected in Table 1 we should have 
$\simeq 55$ GRBs in the same table. Because 
the brightness bias exists, $[(55-9)/55]\times 100 = 84$\% of GRBs should 
have unseen AGs.  Hence, the brightness  bias
alone is able to explain even a ratio $55:9 \simeq 6:1$ for the
observationally dark--non-dark population. This is much bigger than 
the observed  $\simeq 2:1$ ratio.

Of course, there is a great uncertainty concerning the obtained
$\simeq 6:1$ ratio, because of the extrapolation of a property, obtained 
from a sample containing 9 objects to a sample for 211 objects. Also
the cut at $P_{256} = 10${\em ph/(cm$^2$s)} is ad hoc. In addition,
the newer instruments (BeppoSAX, HETE2, INTEGRAL) have 
different thresholds in the gamma-ray band than that of BATSE.
Nevertheless, keeping all this in mind, we stress that
the brightness bias is so strong that even alone it is able to
explain the fact that the two-third fraction of long GRBs have
existing but unseen AGs.

\section{GRBs at very high redshifts and the extinction}

Several papers (\cite{ref:mm95}, 
\cite{ref:mm96}, \cite{ref:hmm96}, \cite{ref:rm97}, 
\cite{ref:lr00}, \cite{ref:sch}, \cite{ref:bag}) estimate that a few 
tens of \% of GRBs are at $5 < z < 20$, where $z$ is the redshift.
In \cite{ref:lin} it is claimed that even the majority of GRBs are at $z 
>10$. It is already an observational evidence (see \cite{ref:klo} for more 
details and references therein) that some dark GRBs are dark due to the 
extinction at the hosts.
Trivially, the existing population of GRBs at very high redshifts 
together with the confirmed extinction strengthen the fraction of the
observationally dark GRBs. 

\section{Conclusion}

The three effects I-III together may well cause that GRBs, detected only in
gamma-ray band but having existing unseen AGs, may have a six or  
more times bigger population than that of GRBs with seen AGs. Hence, {\bf 
dark GRBs in Greiner's list may well be explained by these three 
instrumental  effects alone, and there is no need to introduce any 
intrinsically dark subgroup 
of long GRBs without AGs.} 

\acknowledgments

Thanks are due to the valuable discussions with Claes-Ingvar Bj\"ornsson,
Istv\'an Csa\-bai, Claes Fransson, Johan Peter Uldall Fynbo, Peter 
M\'esz\'aros, G\'abor Tusn\'ady, and Roland Vavrek. The useful remarks of 
the anonymous referee are kindly acknowledged. 
This research was supported through OTKA grants T034549, T48870,
and by a grant from the Wenner-Gren Foundations (A.M.).


\begin{thebibliography}{0}
\bibitem{ref:gre} \BY{Greiner~J.}
  http://www.mpe.mpg.de/$^{\sim}$jcg/grbgen.html.
\bibitem{ref:jak} \BY{Jakobsson~P. et al.} \IN{ApJ} {617}{2004}{L21}.
\bibitem{ref:fyn} \BY{Fynbo~J.U. et al.} \IN{A\&A} {369}{2001}{373}.
\bibitem{ref:stra} \BY{Stratta~G. et al.} \IN{ApJ} {608}{2004}{846}.
\bibitem{ref:lamb} \BY{Lamb D.Q. et al.} \IN{NewA.Rev.} {48}{2004}{423}.
\bibitem{ref:bal} \BY{Bal\'azs~L.G. et al.} \IN{A\&A} {401}{2003}{129}.
\bibitem{ref:bor} \BY{Borgonovo~L.} \IN{A\&A}{418}{2004}{487}.  
\bibitem{ref:bag} \BY{Bagoly~Z. et al.}
  \IN{A\&A}{398}{2003}{919}.
\bibitem{ref:me04} \BY{M\'esz\'aros~ A. et al.} in
\TITLE{Third Rome Workshop on Gamma-Ray Bursts in the Afterglow Era},
edited by \NAME{Feroci M., Frontera F., Masetti N. \atque Piro L.} 
(ASP, San Francisco) 2004, 118.
\bibitem{ref:mee} \BY{Meegan C.A. et al.} 
http://www.gammaray.msfc.nasa.gov/batse.
\bibitem{ref:mm95} \BY{M\'esz\'aros P. \atque  M\'esz\'aros A.} \IN{ApJ}
{449} {1995} {9}.
\bibitem{ref:mm96} \BY{M\'esz\'aros A. \atque  M\'esz\'aros P.} \IN{ApJ} 
{466} {1996} {29}.
\bibitem{ref:hmm96} \BY{Horv\'ath I., M\'esz\'aros, P. \atque 
M\'esz\'aros A.} \IN{ApJ} {470} {1996} {56}.
\bibitem{ref:rm97} \BY{Reichart D.E. \atque M\'esz\'aros P.} 
\IN{ApJ} {483}{1997}{597}.
\bibitem{ref:lr00} \BY{Lamb D.Q. \atque  Reichart D.E.} \IN{ApJ} {536} 
{2000}{1}.
\bibitem{ref:sch} \BY{Schmidt M.} \IN{ApJ} {552} {2000} {36}.
\bibitem{ref:lin} \BY{Lin J.R., Zhang S.N. \atque Li T.P.} \IN{ApJ}
{605} {2004} {579}.
\bibitem{ref:klo} \BY{Klose S. et al.} \IN{ApJ} {592} {2003} {1025}.
\end{thebibliography}
\end{document}